\documentclass[aip,reprint]{revtex4-1}

\usepackage{amsmath}
\usepackage{amssymb}

\usepackage[colorlinks,citecolor=blue,linkcolor=blue]{hyperref}
\usepackage{graphicx}% Include figure files
\usepackage{dcolumn}% Align table columns on decimal point
\usepackage{bm}% bold math

\usepackage{txfonts}

\begin{document}

\title{Current-driven skyrmion dynamics in disordered films}

\author{Joo-Von Kim}
\email{joo-von.kim@c2n.upsaclay.fr}
\affiliation{Centre de Nanosciences et de Nanotechnologies, CNRS, Univ. Paris-Sud, Universit{\'e} Paris-Saclay, 91405 Orsay, France}
\author{Myoung-Woo Yoo}
\affiliation{Centre de Nanosciences et de Nanotechnologies, CNRS, Univ. Paris-Sud, Universit{\'e} Paris-Saclay, 91405 Orsay, France}
\affiliation{Unit{\'e} Mixte de Physique, CNRS, Thales, Univ. Paris-Sud, Universit{\'e} Paris-Saclay, 91767 Palaiseau, France}

\date{\today}

\begin{abstract}
A theoretical study of the current-driven dynamics of magnetic skyrmions in disordered perpendicularly-magnetized ultrathin films is presented. The disorder is simulated as a granular structure in which the local anisotropy varies randomly from grain to grain. The skyrmion velocity is computed for different disorder parameters and ensembles. Similar behavior is seen for spin-torques due to in-plane currents and the spin Hall effect, where a pinning regime can be identified at low currents with a transition towards the disorder-free case at higher currents, similar to domain wall motion in disordered films. Moreover, a current-dependent skyrmion Hall effect and fluctuations in the core radius are found, which result from the interaction with the pinning potential.\end{abstract}

\maketitle

%%
%	begin text
%%

Magnetic skyrmions are nanoscale spin configurations with a nontrivial topology.~\cite{Bogdanov:2001hr, Heinze:2011ic, Romming:2013iq} In ultrathin ferromagnets in contact with a strong spin-orbit material, they are stabilized by interfacial chiral interactions of the Dzyaloshinskii-Moriya form (DMI)~\cite{Dupe:2014fc, Dupe:2016bv} and possess core sizes down to the nanometer range.~\cite{Romming:2015il} Skyrmions can be moved by spin currents and their dynamics depends on their topological properties.~\cite{Komineas:2015bi, Yamane:2016ha} They have been touted as promising candidates for various spintronics applications, such as racetrack memories~\cite{Fert:2013fq, Tomasello:2014ka} and microwave detectors,~\cite{Finocchio:2015cm} and offer possible advantages over domain-wall--based systems because they are less susceptible to certain defects.~\cite{Fert:2013fq, Sampaio:2013kn, Iwasaki:2013hb} Recent experiments have confirmed the existence of room-temperature skyrmions in sputtered multilayer systems,~\cite{Jiang:2015cs, MoreauLuchaire:2016em, Boulle:2016jt, Woo:2016jw, Hrabec:2016wy, Legrand:2017} which is an important milestone toward realizing skyrmion-based devices.

For interface-driven DMI, most material systems investigated to date involve ultrathin ferromagnets with perpendicular magnetic anisotropy. These systems have also been studied extensively for magnetic domain wall dynamics,~\cite{Metaxas:2007fl, Miron:2011fn, Ryu:2013dl, Emori:2013cl, Hrabec:2014hm, Torrejon:2014dr} where observations of strong pinning are common. In the context of skyrmion dynamics, it is natural to enquire whether the same disorder that leads to wall pinning can also have an influence on skyrmion propagation. Experimentally, it has been established that pinning can be strong.~\cite{Hanneken:2016dd} While the effects of boundary edges on skyrmion propagation have been studied,~\cite{Fert:2013fq, Sampaio:2013kn} studies of the role of random disorder to date have been limited to atomistic~\cite{Iwasaki:2013hb, Muller:2015fu} or particle-based models.~\cite{Lin:2013gj, Reichhardt:2015kt, Reichhardt:2016ge}

Here, we revisit the problem by using micromagnetics simulations to model more realistic disorder that is relevant to ultrathin films. Specifically, the disorder is modeled as local variations in the perpendicular anisotropy and the current-driven motion due to current-in-plane and spin Hall effect torques are considered. We find significant pinning at low applied currents and additional contributions to the skyrmion Hall effect that arises from interaction with the disorder potential, which is consistent with previous studies.~\cite{Muller:2015fu, Reichhardt:2016ge} We also examine how the skyrmion core is deformed as it traverses the disorder potential.

We used the MuMax3 code~\cite{Vansteenkiste:2014et} for the micromagnetics simulations. The code integrates numerically the Landau-Lifshitz equation with Gilbert damping and spin torques,
\begin{equation}
\frac{d \mathbf{m}}{dt} = -|\gamma| \mu_0 \mathbf{m} \times \mathbf{H}_{\rm eff} + \alpha \mathbf{m} \times \frac{d \mathbf{m}}{dt} + \mathbf{\Gamma}_{\rm ST},
\end{equation}
where $\mathbf{m} = \mathbf{m}(\mathbf{x},t)$ is a unit vector representing the magnetization state, $\mathbf{H}_{\rm eff}$ is the effective field, $\gamma$ is the gyromagnetic constant, $\mu_0$ is the permeability of free space, $\alpha$ is the damping constant, and $\mathbf{\Gamma}_{\rm ST}$ represents spin torques. We model a perpendicularly-magnetized ferromagnetic film with a thickness of $d = 0.6$ nm. In Cartesian coordinates, the $xy$ plane represents the film plane and $z$ is the direction of the uniaxial anisotropy, perpendicular to the film plane. We assumed an exchange constant of $A = 16$ pJ/m, a uniaxial anisotropy of $K_u = 1.3$ MJ/m$^3$, and a saturation magnetization of $M_s = 1.1$ MA/m. These parameters are consistent with the values obtained for ultrathin Co in Pt/Co/AlOx multilayers,~\cite{Belmeguenai:2015hj} which exhibit a strong DMI. For the dynamics we assumed $\alpha = 0.3$, which is consistent with recent experiments.~\cite{Schellekens:2013it} In order to simulate disorder, we considered a random grain structure in which the local anisotropy of each grain $i$, $K_{u,i}$, is drawn from a Gaussian distribution centered on the mean value $K_u$ with a standard deviation of $\delta K$.~\cite{Voto:2016kr, GarciaSanchez:2016cx} The grain structure is constructed using Voronoi tessellation and different average grain sizes, $\langle L \rangle$, are considered.

\begin{figure}
\includegraphics[width=8.5cm]{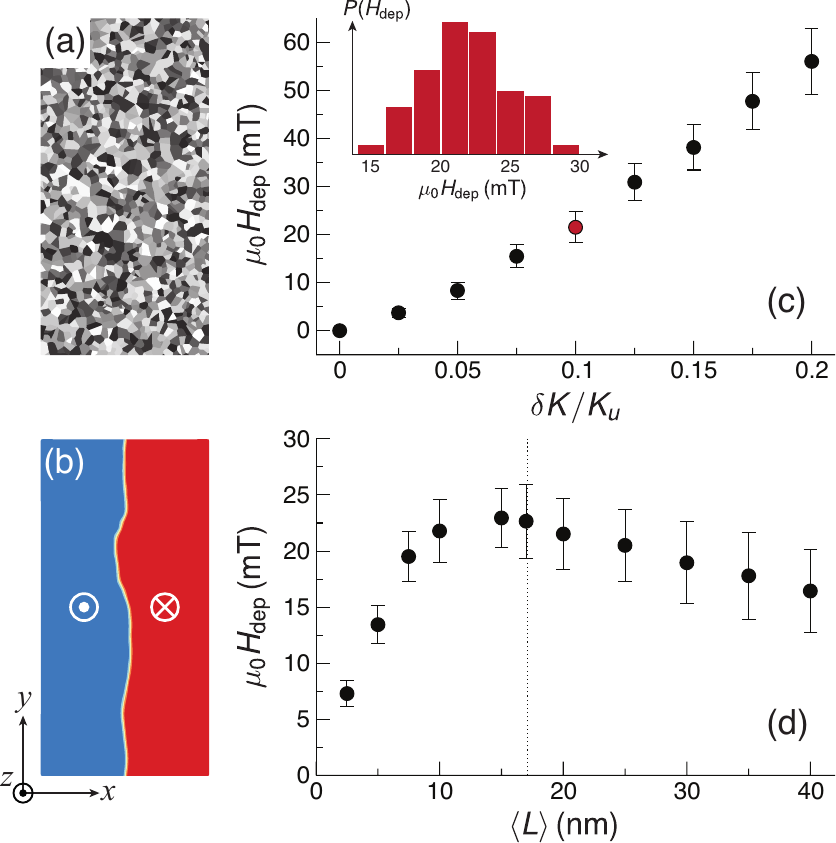}
\caption{Characteristics of the simulated disorder. (a) Example of grain distribution for $ \langle L \rangle =20$ nm. The image represents a simulated region of 0.5 $\mu$m $\times$ 1.0 $\mu$m. (b) Pinned domain wall with the disorder in (a). (c) Domain wall depinning field, $\mu_0 H_{\rm dep}$, as a function of the strength of the anisotropy fluctuations, $\delta K / K_u$, for $ \langle L \rangle =20$ nm. The inset shows the probability distribution of $H_{\rm dep}$ for $\delta K / K_u = 0.1$.  (d) Depinning field as a function of average grain size for $\delta K / K_u = 0.1$. The dashed line indicates the wall width, $\pi \Delta \approx 17.1$ nm. 
}
\label{fig:depin}
\end{figure}

Before examining the skyrmion dynamics in detail, we first discuss how such disorder can be related to observable quantities in experiment. A direct means to characterize the disorder is through the depinning field, $H_{\rm dep}$, for magnetic domain wall propagation. This is defined as the threshold field at which a magnetic domain wall can propagate freely without being impeded by any pinning (and corresponds to the situation in which the force acting on the wall is the same sign everywhere). It is a quantity that is readily accessible experimentally for continuous films, so it represents a useful measure of the disorder to which simulations can be benchmarked.  Results of micromagnetics simulations of $H_{\rm dep}$ are shown in Fig.~\ref{fig:depin}.
The simulations are performed for a system with lateral dimensions of 0.5 $\mu$m $\times$ 1.0 $\mu$m that is discretized using 512$\times$1024$\times$1 finite difference cells (the magnetization is assumed to be uniform across the film thickness). An example of the grain structure with $\langle L \rangle = 20$ nm is given in Fig.~\ref{fig:depin}(a). The initial micromagnetic state consists of two domains, one oriented along $+z$ for $x<0$ and the other along $-z$ for $x>0$, which results in a domain wall running across the width ($y$) of the system. Periodic boundary conditions are applied along $y$ to avoid edge effects. The initial state is then relaxed using energy minimization, which results in a rugged domain wall structure [Fig.~\ref{fig:depin}(b)]. Next, a magnetic field is applied along $+z$ and increased incrementally, where at each field step the equilibrium configuration is found using energy minimization. This proceeds until the domain wall is completely depinned and sweeps across the system, resulting in a uniformly magnetized state along $+z$. We designate $H_{\rm dep}$ as the field at which this occurs. For each set of $\delta K / K_u$ and $ \langle L \rangle$, the simulations are repeated for 100 different realizations of the disorder so that ensemble averages can be obtained.

The dependence of the depinning field on the anisotropy variation is shown in Fig.~\ref{fig:depin}(c). A monotonic increase is seen, which can be expected since larger variations in the anisotropy are likely to result in larger spatial variations in the energy landscape and therefore stronger pinning. The error bars correspond to one standard deviation of the pinning field distribution, where the distribution for $\delta K / K_u = 0.1$ is given in the inset of Fig.~\ref{fig:depin}(c). In Fig.~\ref{fig:depin}(d), the variation of $H_{\rm dep}$ with the average grain size (for $\delta K / K_u = 0.1$) is shown. Unlike for anisotropy fluctuations, the variation is non-monotonic and exhibits a peak around $\langle L \rangle \simeq 15$ nm, which is preceded by a rapid increase and followed by a slower decrease as  $\langle L \rangle$ is increased. On this plot, the domain wall width, $\pi \Delta = \pi \sqrt{A/K_0}$, where $K_0 = K_u - \mu_0 M_s^2/2$ is the effective perpendicular anisotropy, is shown and roughly coincides with the position of the maximum in the pinning field. We can understand this result as follows. For sufficiently small grains, $\langle L \rangle \ll \pi \Delta$, the anisotropy variations are averaged out over the wall width and therefore the pinning potential is smoothed out, leading to low pinning fields. At the opposite limit, $\langle L \rangle \gg \pi \Delta$, the domain wall traverses larger regions in which the anisotropy remains constant, so the pinning field in that case is predominantly determined by step changes in the anisotropy at grain boundaries. Note that $\Delta$ is relevant to skyrmions, since it is also characteristic scale of the double-soliton \emph{ansatz}~\cite{Braun:1994ff} that describes the skyrmion core profile.~\cite{Romming:2015il}

We now discuss the role of anisotropy disorder on current-driven skyrmion dynamics. We model a system with dimensions of 0.5 $\mu$m $\times$ 0.5 $\mu$m $\times$ 0.6 nm that is discretized with 512$\times$512$\times$1 finite difference cells. Periodic boundary conditions are assumed along $x$ and $y$, which mimics an infinite system and avoids edge boundary effects.~\cite{Yoo:2017} The skyrmion Hall effect ensures that the motion is not parallel to the simulation grid ($x$ or $y$), which allows different grains to be traversed as the skyrmion wraps around the simulation grid. We assume a DMI constant of $D = 2.7$ mJ/m$^2$, which is consistent with experimental results.~\cite{Belmeguenai:2015hj}

The simulations are performed as follows. For each realization of the disorder, the initial configuration comprises a single N{\'e}el-type skyrmion at the center of the simulation grid, with the core magnetization oriented toward $+z$ and the uniform background magnetization along $-z$. The simulation is run for 10 ns at a given current density and the skyrmion displacement is computed over this interval. This is performed for 50 different realizations of the disorder for each applied current density. While no thermal fluctuations are taken into account (to minimize computation time), the simulations mimic experiments in which observations are only made after successive current pulses are applied. Moreover, thermal activation in real systems would result in different starting points prior to each pulse, which is accounted for here with our ensemble averaging of the disorder. We note that skyrmion annihilation occurs under certain conditions (large currents, strong disorder); for such cases, the averages are performed only for the duration for which the skyrmion is present.

\begin{figure}
\includegraphics[width=8.5cm]{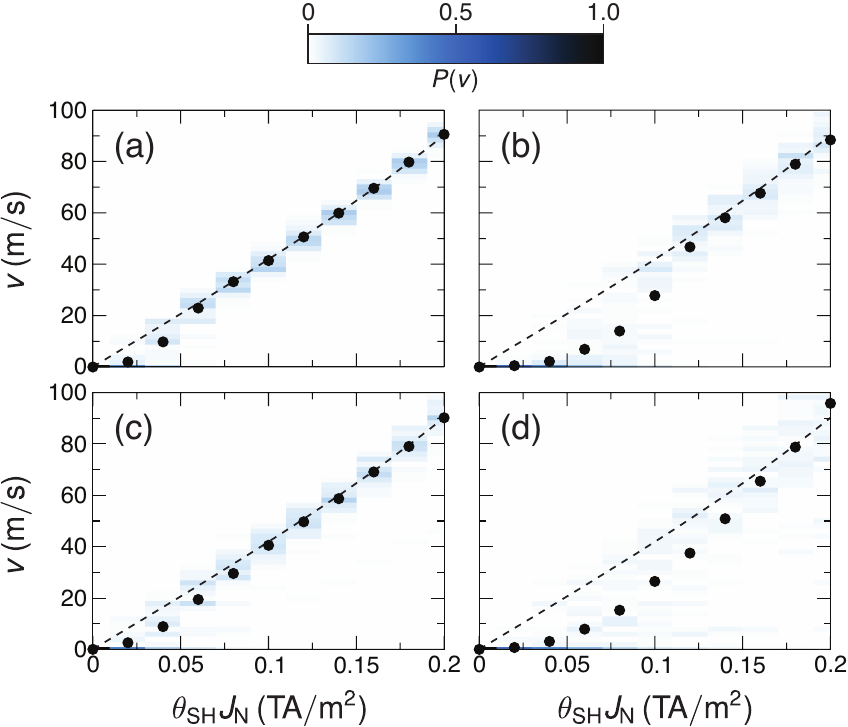}
\caption{Average skyrmion velocity, $v$, as a function applied spin Hall current density, $\theta_{\rm SH} J_{\rm N}$, for different disorder parameters: (a) $\delta K/K_{u} = 0.05$, $\langle L \rangle = 10$ nm; (b) $\delta K/K_{u} = 0.1$, $\langle L \rangle = 10$ nm; (c) $\delta K/K_{u} = 0.05$, $\langle L \rangle = 20$ nm; (d) $\delta K/K_{u} = 0.1$, $\langle L \rangle = 20$ nm. Points are simulation data and dashed lines correspond to the behavior in the disorder-free system. The background of each plot is a probability density map of the velocity, with the legend given at the top of the figure.}
\label{fig:vel_vs_JSHE}
\end{figure}

In Fig.~\ref{fig:vel_vs_JSHE}, we present the average skyrmion velocity as a function of current-induced spin Hall torques for different disorder parameters.
We assume a hypothetical current $J_{\rm N} \hat{\mathbf{x}}$ flows in an adjacent heavy-metal buffer layer, which generates a spin current polarized along $\hat{\mathbf{y}}$ that flows along $\hat{\mathbf{z}}$ into the ferromagnet. $\theta_{\rm SH}$ is the spin Hall angle of the heavy metal layer. This leads to a torque of the form
\begin{equation}
\mathbf{\Gamma}_{\rm ST,SH} = \frac{\hbar \gamma}{2 e M_s d} \theta_{\rm SH} J_{\rm N} \, \mathbf{m} \times \left( \mathbf{m} \times \hat{\mathbf{y}} \right),
\end{equation}
where $e$ is the electron charge. For the disorder parameters considered, we can identify a pinning regime at low current densities in which the average velocity is significantly below the value for the disorder-free case. This behavior is mainly determined by events for which the skyrmion becomes pinned by the disorder within the 10-ns simulation window, either at the onset or after a certain duration during which a finite displacement takes place. We present the probability density of the velocities as a color map, where significant spread can be seen for stronger disorder, as expected. We note that the exponential-like increase at low currents, followed by a smooth transition toward the disorder-free case, is typical of driven interface motion in disordered media,~\cite{Chauve:2000jt} such as domain wall propagation in perpendicular anisotropy materials.~\cite{Metaxas:2007fl} Our results therefore highlight the similarity with the depinning dynamics of domain walls and confirm that skyrmions are not impervious to defect-induced pinning for realistic disorder. We note that similar trends have been observed in recent experiments on current-driven skyrmion motion.~\cite{Woo:2016jw, Hrabec:2016wy, Legrand:2017}

\begin{figure}
\includegraphics[width=8.5cm]{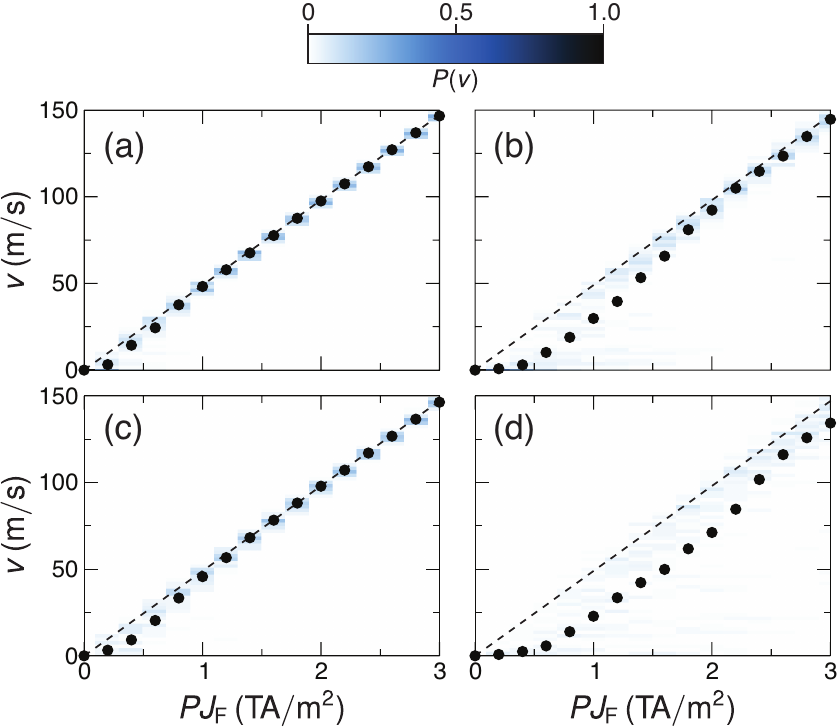}
\caption{Average skyrmion velocity, $v$, as a function applied in-plane current density, $P J_{\rm F}$, for different disorder parameters: (a) $\delta K/K_{u} = 0.05$, $\langle L \rangle = 10$ nm; (b) $\delta K/K_{u} = 0.1$, $\langle L \rangle = 10$ nm; (c) $\delta K/K_{u} = 0.05$, $\langle L \rangle = 20$ nm; (d) $\delta K/K_{u} = 0.1$, $\langle L \rangle = 20$ nm. Points are simulation data and dashed lines correspond to the behavior in the disorder-free system. The background of each plot is a probability density map of the velocity, with the legend given at the top of the figure.}
\label{fig:vel_vs_JCIP}
\end{figure}

In Fig.~\ref{fig:vel_vs_JCIP}, we present the average skyrmion velocity as a function of in-plane applied currents.
Here, the spin torques are related to the in-plane (CIP) flow of spin-polarized currents across magnetic textures, which involves adiabatic and nonadiabatic contributions,~\cite{Zhang:2004hs} 
\begin{equation}
\mathbf{\Gamma}_{\rm ST,CIP} = -\mathbf{u} \cdot \nabla \mathbf{m} + \beta \, \mathbf{m} \times \left( \mathbf{u} \cdot \nabla \mathbf{m} \right),
\end{equation}
where $\mathbf{u} = (\hbar \gamma/2e M_s) P \,  \mathbf{J}_{\rm F} $ is an effective spin drift velocity associated with the in-plane current (density) flowing through the ferromagnet, $J_{\rm F}$.  We neglected the nonadiabatic term ($\beta = 0$) for the simulation results shown in Fig.~\ref{fig:vel_vs_JCIP} as it only affects the intrinsic skyrmion Hall angle (under the rigid core assumption). Nevertheless, we have performed simulations to verify that a nonzero $\beta$ term does not change our findings. The torques are determined by the current density $J_{\rm F} \hat{\mathbf{x}}$ flowing through the ferromagnet with a spin polarization $P$. To facilitate comparisons between the spin Hall torques and the CIP torques, we present results for current densities that result in a similar range of average velocities. Besides the obvious difference in spin torque efficiency, we note that the behavior is qualitatively similar for CIP torques where a transition between the pinning and disorder-free regimes can be seen.

\begin{figure}
\includegraphics[width=8cm]{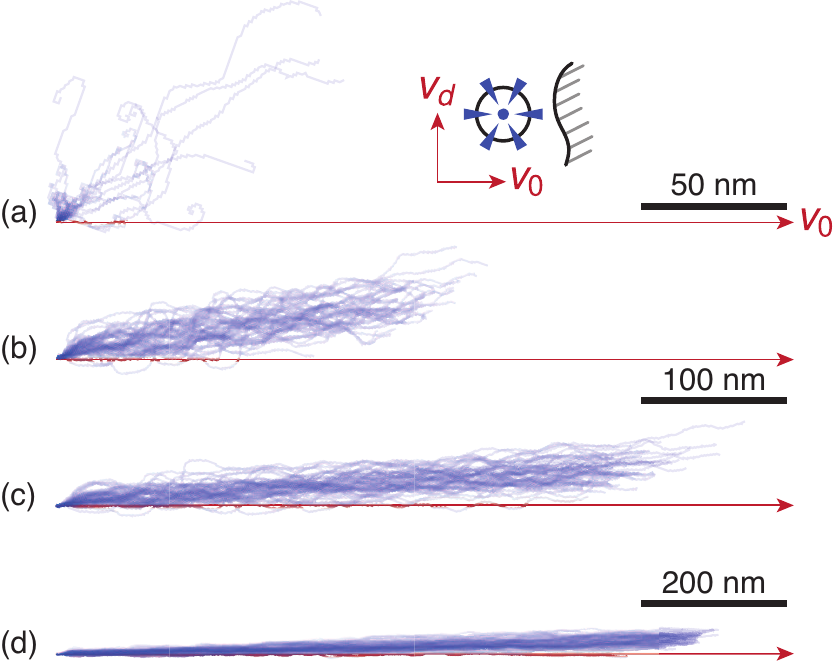}
\caption{Skyrmion trajectories with disorder parameters of $\delta K/K_{u} = 0.05$ and $\langle L \rangle = 10$ nm under different $\theta_{\rm SH} J_{\rm N}$: (a) 0.02 TA/m$^2$, (b) 0.06 TA/m$^2$, (c) 0.1 TA/m$^2$, and (d) 0.2 TA/m$^2$. The horizontal axis ($v_0$) indicates the propagation direction in the disorder-free case. The length scales are given in each sub-figure, where for (b) and (c) the scale bar of 100 nm applies. The inset in (a) illustrates the extrinsic skyrmion Hall motion along $v_d$ due to a potential barrier.}
\label{fig:trajectories}
\end{figure}

The disorder potential strongly influences the direction of the skyrmion propagation. In Fig.~\ref{fig:trajectories}, the trajectories for 50 different disorder realizations ($\delta K/K_{u} = 0.05$, $\langle L \rangle = 10$) are shown for four values of $\theta_{\rm SH} J_{\rm N}$.
The trajectories are presented with the same initial position and relative to the propagation direction of the disorder-free case, which is along the horizontal axis denoted by $v_0$. For the lowest current shown [Fig.~\ref{fig:trajectories}(a)], we observe that most realizations lead to a pinned skyrmion close to its initial position, while only few cases of propagation over tens of nm are seen. We can also observe spirals in some of the trajectories, which possess the same handedness and results from the gyrotropic nature of the skyrmion motion. This is consistent with previous studies on skyrmion pinning.~\cite{Liu:2013in} We note that the trajectories do not lead to an average displacement along $v_0$, but rather at an appreciable angle [approximately 45$^\circ$ in Fig.~\ref{fig:trajectories}(a)].  This result can be understood in terms of the gyrotropic response of a skyrmion to a force; as the skyrmion is driven toward a boundary edge (in our case, a defect-induced potential barrier), the restoring force due to this edge drives the skyrmion along a specific direction perpendicular to this force. As such, each time the skyrmion encounters a potential barrier whilst propagating along the $v_0$ direction, it experiences a restoring force along $-v_0$ that results in an addition deflection along $v_d$, perpendicular to $v_0$. This extrinsic Hall motion becomes less pronounced as the current is increased [Figs.~\ref{fig:trajectories}(b)-(d)], where the current-driven torques become sufficiently large to overcome the pinning potential. Note, however, that in most cases the trajectories exhibiting the largest displacements also exhibit the strongest deflection, which suggests that the skyrmion can circumvent strong potential barriers through the additional Hall motion. These results agree with an analytical approach in which the deflection is predicted within the rigid core approximation~\cite{Muller:2015fu} and particle simulations.~\cite{Reichhardt:2016ge}

\begin{figure}
\includegraphics[width=8.5cm]{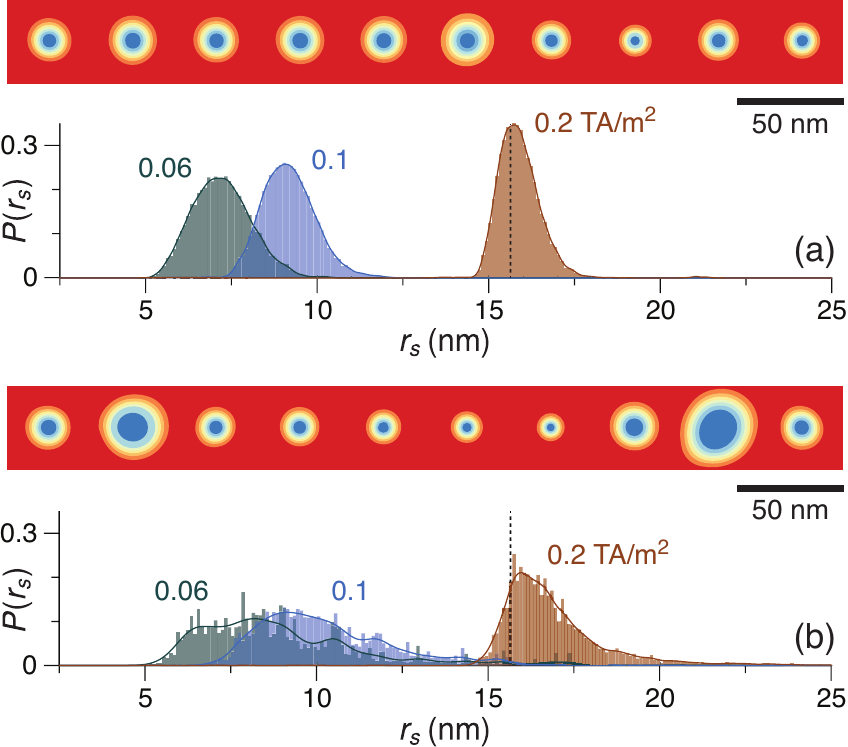}
\caption{Distribution of the effective skyrmion core radius, $r_s$, at different applied current densities, $\theta_{\rm SH} J_{\rm N}$, with disorder parameters of (a) $\delta K/K_{u} = 0.05$, $\langle L \rangle = 10$ nm and (b) $\delta K/K_{u} = 0.1$, $\langle L \rangle = 20$ nm. The dashed line indicates the equilibrium skyrmion radius in the disorder-free case. The insets in (a,b) show a sequence of snapshots of the skyrmion core profile, taken at 1 ns intervals, for a sample trajectory under $\theta_{\rm SH} J_{\rm N} = 0.1$ TA/m$^2$.}
\label{fig:size}
\end{figure}

The skyrmion core fluctuates in size as it traverses the disorder potential. In Fig.~\ref{fig:size}, we present the distribution of the average core radius, $r_s$, for different applied spin Hall currents and disorder parameters, along with snapshots of the core profile.
$r_s$ is computed using the double-soliton ansatz for the $m_z$ component,~\cite{Braun:1994ff, Romming:2015il, GarciaSanchez:2016cx} where we equate the area bound by the curve $m_z = 0$ of the simulated profile with the radius $r_s$ that gives the same area using $m_z(r) = 4 \cosh{\left(c\right)}\left[ \cosh\left( 2c \right) + \cosh\left( 2r /\Delta \right) \right]^{-1}-1$. The distribution is constructed from values taken at 20 ps intervals of each simulation run. We observe that $r_s$ can fluctuate over a range of $\sim 5$ nm as the skyrmion traverses the potential. The mean value depends on the applied current density, where stronger compressive effects are seen at lower currents at which pinning is dominant. This is analogous to the dynamics seen for propagation along boundary edges, where the combination between spin torques and the edge confinement leads to a reduction in the core size.~\cite{GarciaSanchez:2016cx, Yoo:2017} For larger currents at which propagation approaches the disorder-free case, the mean value of $r_s$ is close to its equilibrium value (e.g., 0.2 TA/m$^2$ in Fig.~\ref{fig:size}). Stronger pinning results in a wider distribution of $r_s$, which is accompanied by larger deformations in the core profile [Fig.~\ref{fig:size}(b)].

These results highlight the outstanding challenges for devising reliable skyrmion-based devices. First, pinning due to realistic disorder cannot be neglected and the skyrmion topology alone is insufficient to negate such effects. The disorder parameters considered lead to domain wall pinning fields of tens of mT, which is typical of experimental systems. Second, the disorder leads to an extrinsic skyrmion Hall effect that is current-dependent. Device schemes that require control of the propagation direction should account for this issue.~\cite{Muller:2017hb} Third, the core size fluctuates as the skyrmion propagates, which would lead to additional noise for detection schemes that rely on the surface area of reversed magnetization, such as tunnel magnetoresistance or stray magnetic fields. As these results suggest, optimizing material properties to minimize pinning (including domain walls) would represent a key step toward meeting these challenges. Alternatively, skyrmions in disordered systems could be useful for stochastic-based computing schemes.~\cite{Mizrahi:2016kp, Pinna:2017}

%%
%	end text
%%

%%%
%	Acknowledgements
%%%
\begin{acknowledgments}
The authors acknowledge fruitful discussions with V. Cros, A. Hrabec, J. M{\"u}ller, N. Reyren, S. Rohart, and R. Soucaille. The authors are grateful to A. Thiaville for hosting part of the computational cluster during the downtime of the C2N laboratory. This work was supported by the Agence Nationale de la Recherche (France) under Contract No. ANR-14-CE26-0012 (Ultrasky) and the Horizon2020 Framework Programme of the European Commission (H2020-FETOPEN-2014-2015-RIA) under Contract No. 665095 (MAGicSky).
\end{acknowledgments}

%%
% 		References
%%
\bibliography{articles}

\end{document}